\documentclass[12pt]{article}
\let\section=\subsection     \let\subsection=\subsubsection                
\usepackage{graphicx}

\begin{document}
\begin{center}
   {\large \bf Photoexcitation of Baryon Resonances - News from the
   D$_{13}$(1520)}\\[5mm]
   B. Krusche\\[5mm]
   {\small \it  Department of Physics and Astronomy, University of Basel\\
   CH-4056 Basel, Klingelberstr. 82, Switzerland\\[8mm] }
\end{center}
\begin{abstract}\noindent
The so-called second resonance region of the nucleon comprises the states
P$_{11}$(1440), D$_{13}$(1520) and S$_{11}$(1535). During the last few years
photoproduction experiments have largely contributed to a better understanding
of these states, but the strong suppression of the resonance
structure in total photoabsorption experiments from nuclei is still not
understood. The D$_{13}$-resonance dominates the resonance structure due to its
large photon coupling. In this contribution new results for the excitation and
decay modes of the D$_{13}$ on the free nucleon and first results for the
in-medium behavior of the resonance are summarised.  
\end{abstract}

\section{Introduction}

Due to their different 
couplings to the initial photon - nucleon and the final meson - nucleon states 
the low lying nucleon resonances D$_{13}$(1520) and S$_{11}$(1535) can be 
separated to a large extent: The production of $\eta$-mesons
proceeds almost exclusively via the excitation of the S$_{11}$,
while the largest resonance contributions to single and double pion 
production come from the D$_{13}$. Using this selectivity,
the properties of the resonances, when excited on the free proton
or quasifree neutron, have been studied in much detail during the last 
few years via $\eta$-photoproduction 
\cite{Krusche_1,Krusche_2,Bock,Ajaka,Krusche_3,Hoffmann,Hejny} 
and single and double pion photoproduction reactions 
\cite{Braghieri,Tejedor,Haerter,Zabrodin,Krusche_4,Kleber,Wolf}.
The excellent quality of the data  
allowed precise determinations of the resonance properties, like the extraction
of a 0.05\% - 0.08\% branching ratio for the 
D$_{13}\rightarrow \mbox{N}\eta$ decay \cite{Tiator}. 

Much less is known about the behavior of the isobars 
inside the nuclear medium, where a number of modifications arise. 
The most trivial is the broadening of the excitation functions due to 
Fermi motion. The decay of the resonances is modified by Pauli-blocking
of final states, which reduces the resonance widths, and by additional decay
channels like $\mbox{N}^{\star}\mbox{N}\rightarrow \mbox{NN}$ which cause 
the collisional broadening. Both effects cancel to some extent and 
it is a priori not clear which one will dominate.
A very exciting possibility is that the resonance
widths could be sensitive to in-medium mass modifications of mesons arising 
from chiral restoration effects. The D$_{13}$-resonance for 
example has a 15 - 25\% decay branch to the $\mbox{N}\rho$-channel \cite{PDG}, 
which is only fed from the low energy tail of the $\rho$ mass distribution.
This means that the resonance width is very sensitive to the $\rho$ mass
distribution.
The first experimental investigation of the second resonance region for nuclei
was done with total photoabsorption. The results showed an almost complete 
absence of the resonance bump for $^{4}$He and heavier nuclei 
\cite{Frommhold,Bianchi,MacCormick}, which up to now has not been understood. 

In this talk new experimental results are presented which shed light on the
contribution of the D$_{13}$ resonance to the second resonance bump via the 
double pion decay channels and on the in-medium behavior of the resonance.

\section{Double pion production and the D$_{13}$(1520)}

The cross sections for single meson photoproduction 
(pions and $\eta$-mesons) and double pion photoproduction are almost equal at
incident photon energies between 600 and 800 MeV. 
Moreover, most of the rise of
the total photoabsorption cross section from the dip above the
\begin{figure}[b]
\begin{minipage}{0.cm}
   \includegraphics[width=7.3cm]{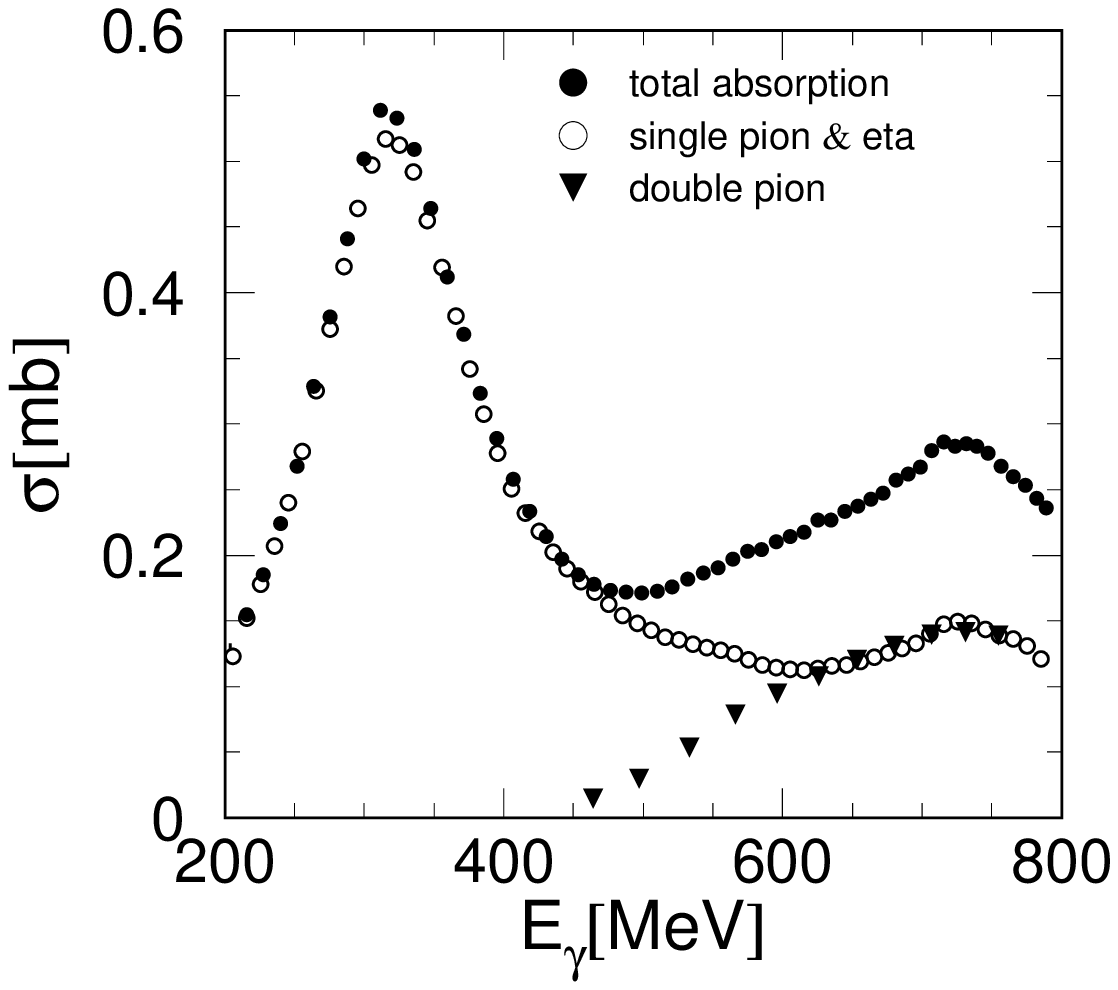}\\
\end{minipage}
\hspace{7.2cm}
\begin{minipage}{0.cm}
   \includegraphics[width=6.5cm]{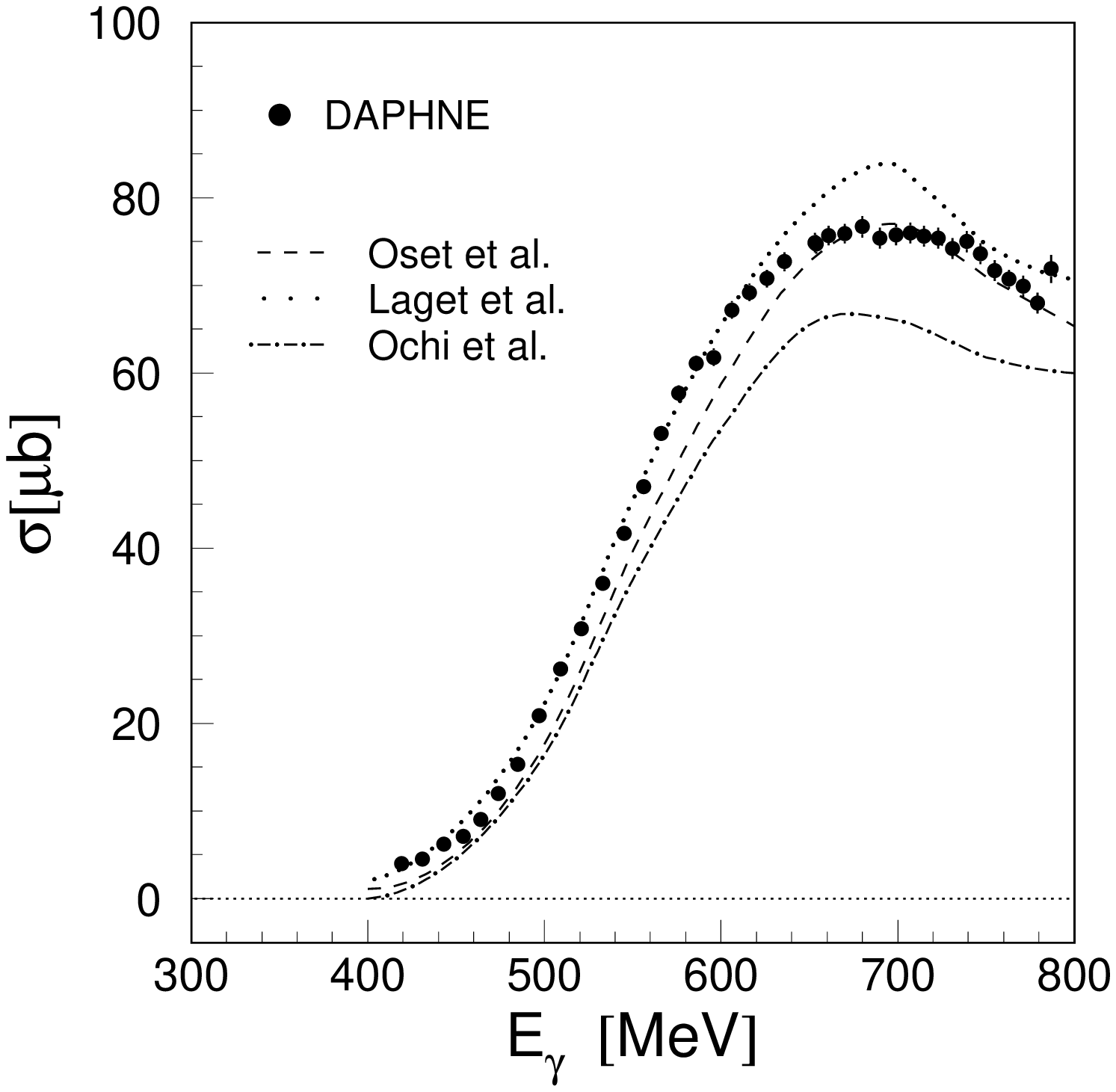}\\
\end{minipage}
\vspace*{-0.8cm}
\begin{center} \parbox{14cm}{
 \caption[]{\label{fig:1}
Left hand side: decomposition of the total photoabsorption cross section of 
the proton into single and double meson production. Right hand side: total 
cross section for $\gamma p\rightarrow p\pi^+\pi^-$ \cite{Braghieri}
compared to models \cite{Gomez,Murphy,Ochi}.
}} 
 \end{center}
\end{figure}
$\Delta$-resonance to the peak of the second resonance bump is due to double
pion production (see fig.\ref{fig:1}, left hand side).
Any detailed interpretation of the second resonance bump therefore requires the
understanding of double pion production. 
Previously it was not even clear which role the P$_{11}$-, D$_{13}$- and 
S$_{11}$-resonances play for double pion production.
Background terms like the $\Delta$-Kroll-Rudermann (KR) and the 
$\Delta$-pion-pole term, which instead involve the excitation of the $\Delta$, 
are important at least for the charged double pion channels.

Among the possible double pion production reactions previously only
$\gamma p\rightarrow p\pi^+\pi^-$ was measured with good precision. 
The total cross section, which is in reasonable agreement with model 
calculations (see fig.\ref{fig:1}, right hand side), is very small
between threshold at $\approx$310 MeV and $\approx$400 MeV. It rises sharply
from $\approx$400 MeV to a maximum at $\approx$650 MeV. This rise is accompanied
by a strong peak at the mass of the $\Delta$-resonance in the invariant mass
distribution of the $p\pi^+$-pair. This peak is absent in the $p\pi^-$ invariant
mass. A large contribution to the cross section is therefore assigned in all
models to the $\gamma p\rightarrow\Delta^{++}\pi^-$-reaction via the 
$\Delta$-KR and the $\Delta$-pion pole terms. 
\begin{figure}[h]
\begin{minipage}{0.cm}
   \includegraphics[width=7.cm]{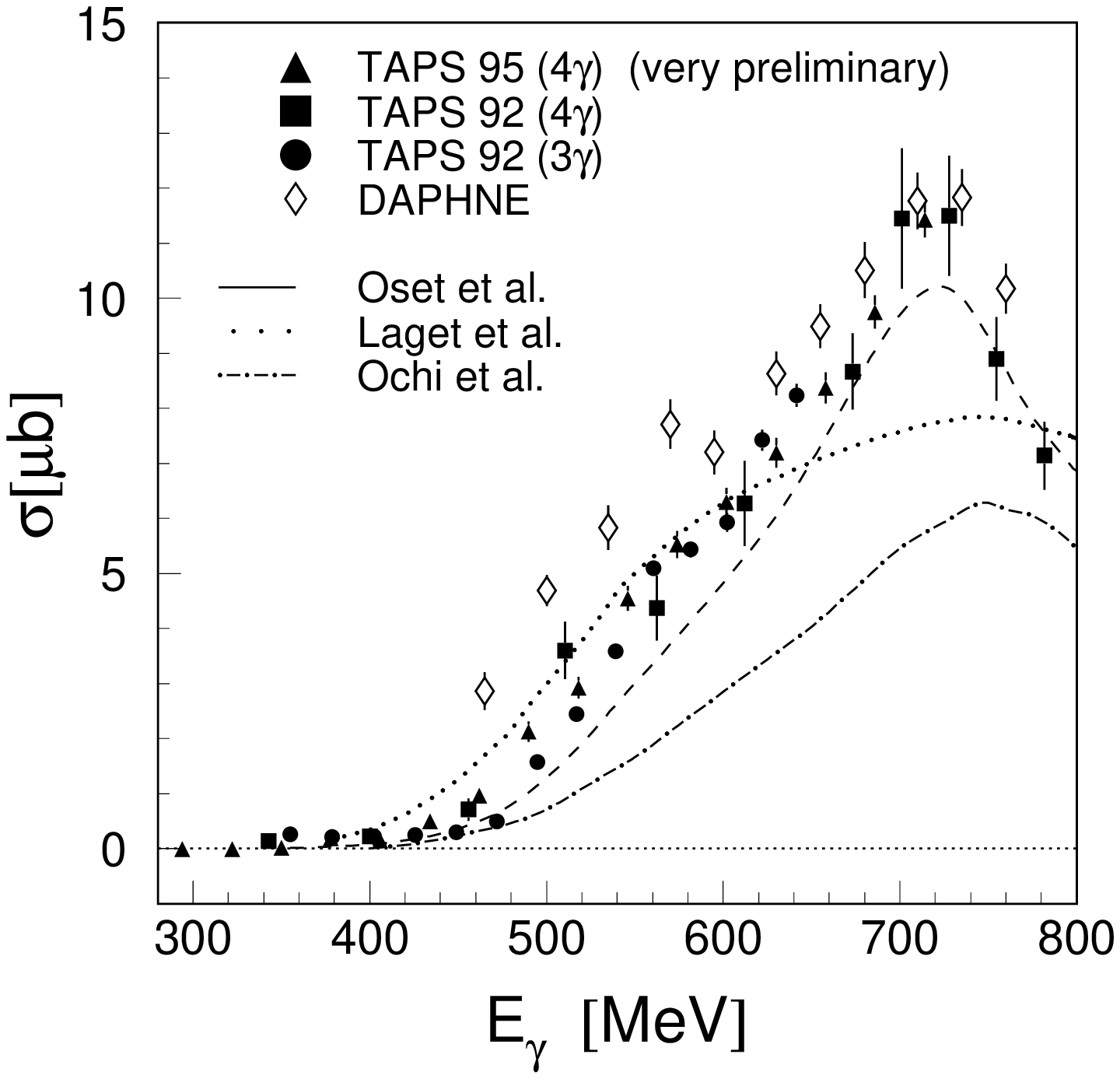}\\
\end{minipage}
\hspace{7.cm}
\begin{minipage}{0.cm}
   \includegraphics[width=7.cm]{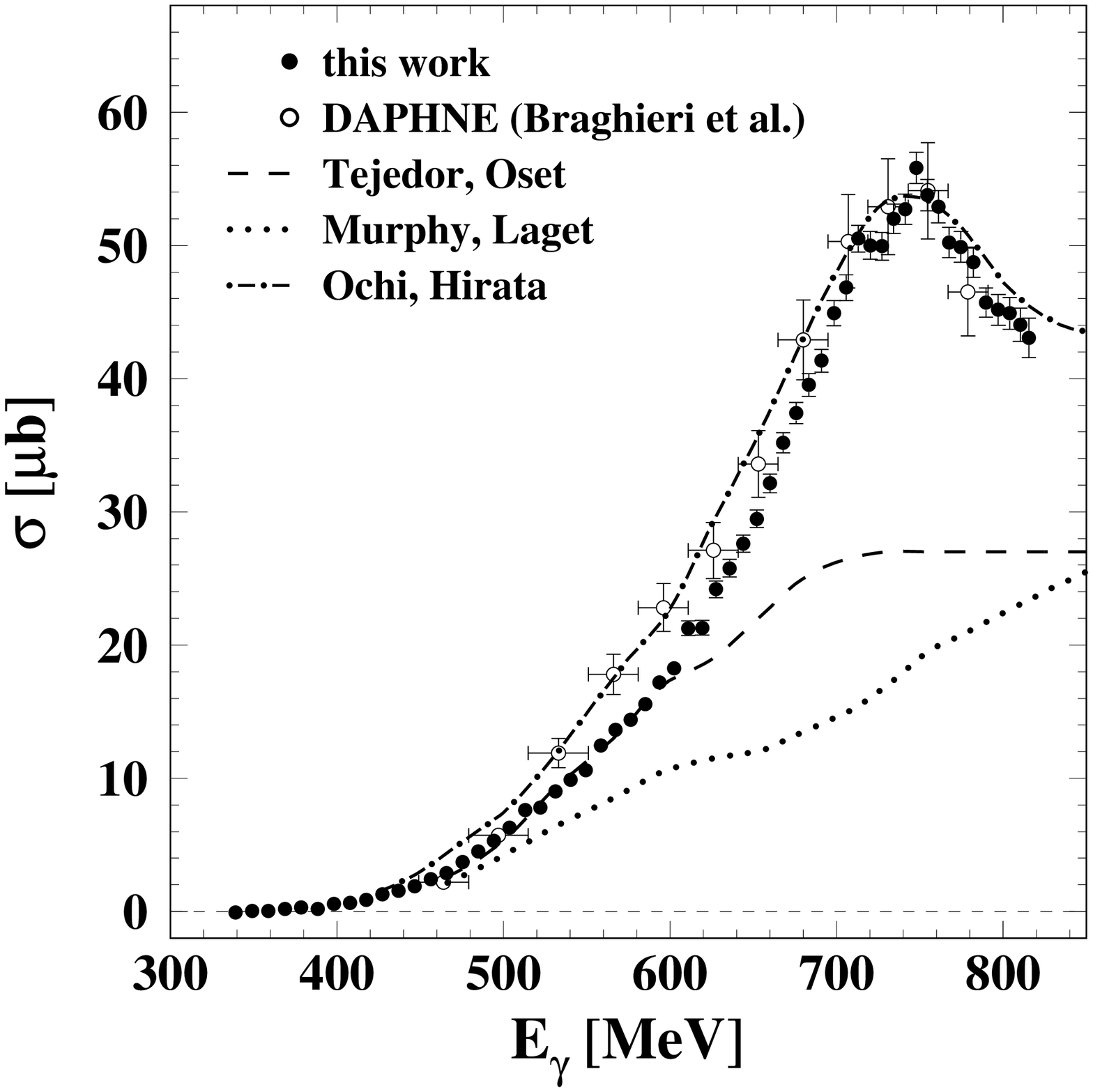}\\
\end{minipage}
\vspace*{-0.8cm}
\begin{center} \parbox{14cm}{
 \caption[]{\label{fig:2}
Total cross sections of the $\gamma p\rightarrow p\pi^o\pi^o$ reaction (left
hand side) and the $\gamma p\rightarrow n\pi^+\pi^o$-reactions compared to model
predictions \cite{Gomez,Murphy,Ochi}.
}} 
 \end{center}
\end{figure}

The situation is very different for the final states involving neutral pions.
The total cross sections for $\gamma p\rightarrow p\pi^o\pi^o$
and $\gamma p\rightarrow n\pi^+\pi^o$ measured in Mainz with the TAPS
\cite{Haerter,Wolf,Langgaertner} and DAPHNE \cite{Braghieri} detectors 
are compared in fig.\ref{fig:2} to the model calculations of Gomez-Tejedor 
et al. \cite{Gomez}, Murphy et al.  \cite{Murphy} and Ochi et al. \cite{Ochi}.
None of the models agrees with both data sets.  
The calculation by Ochi et al. \cite{Ochi}, was developed later with special 
emphasis on the $\pi^+\pi^o$ final state. 

In case of the $\pi^o\pi^o$ final state the two models from refs. 
\cite{Gomez,Murphy} made very different predictions.
One of them \cite{Gomez} predicted as dominant process the sequential decay of 
the D$_{13}(1520)$ resonance via a $\Delta\pi$ intermediate state, the other 
\cite{Murphy} the decay of the P$_{11}(1440)$ resonance via a correlated pair 
of pions in a relative s-wave. The total cross section 
(see fig.\ref{fig:2}) is in better agreement with the prediction from 
ref. \cite{Gomez}, but the problem was finally solved by the study of the
the invariant mass distributions \cite{Haerter,Wolf}.
The pion - pion invariant mass distributions are similar to phase space 
behavior, while a strong deviation from phase space was predicted 
for the correlated two pion decay of the P$_{11}$ in \cite{Murphy}. 
The pion - proton invariant mass deviates from phase space 
and peaks at the $\Delta$ mass as expected for a sequential 
$N^{\star}\rightarrow\Delta\pi^o\rightarrow\ N\pi^o\pi^o$ decay and as
predicted in \cite{Gomez}. The high quality invariant mass distributions
available now \cite{Wolf}, will certainly allow a more detailed analysis. 

The first measurement of the $\gamma p\rightarrow\pi^o\pi^+$ reaction 
\cite{Braghieri} came up with a total cross section that was strongly 
underestimated by the predictions from the then available models
\cite{Gomez,Murphy} (see fig.\ref{fig:2}, right hand side).
This experimental result was confirmed by a measurement with 
the TAPS detector \cite{Langgaertner} and a similar situation was found for 
the $\gamma n\rightarrow p\pi^-\pi^o$ reaction \cite{Zabrodin}. Obviously an 
important contribution was missing in the models.

\begin{figure}[t]
\begin{minipage}{0.cm}
   \includegraphics[width=6.5cm]{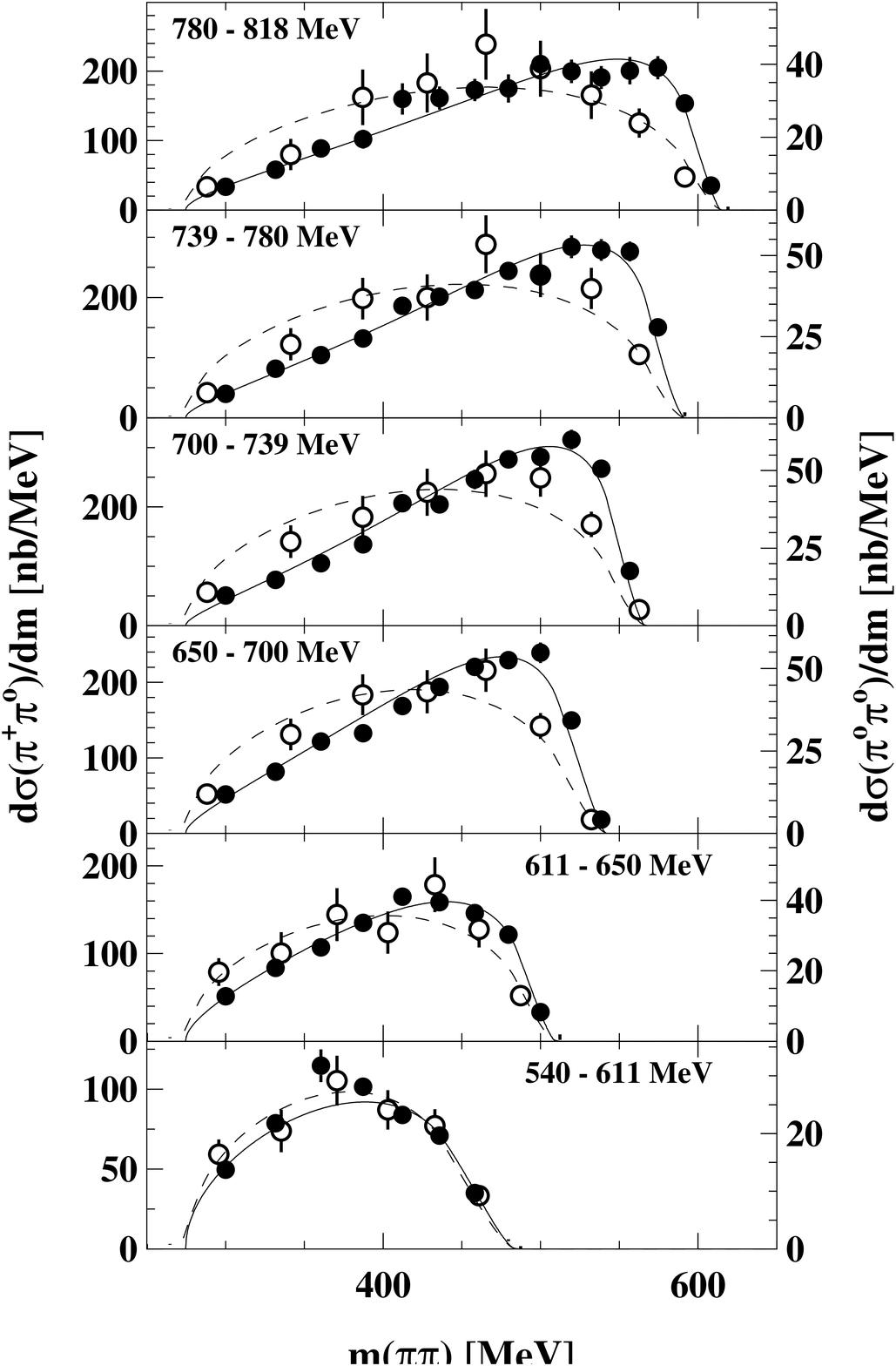}\\
\end{minipage}
\hspace{7.cm}
\begin{minipage}{0.cm}
   \includegraphics[height=9.5cm,width=6.5cm]{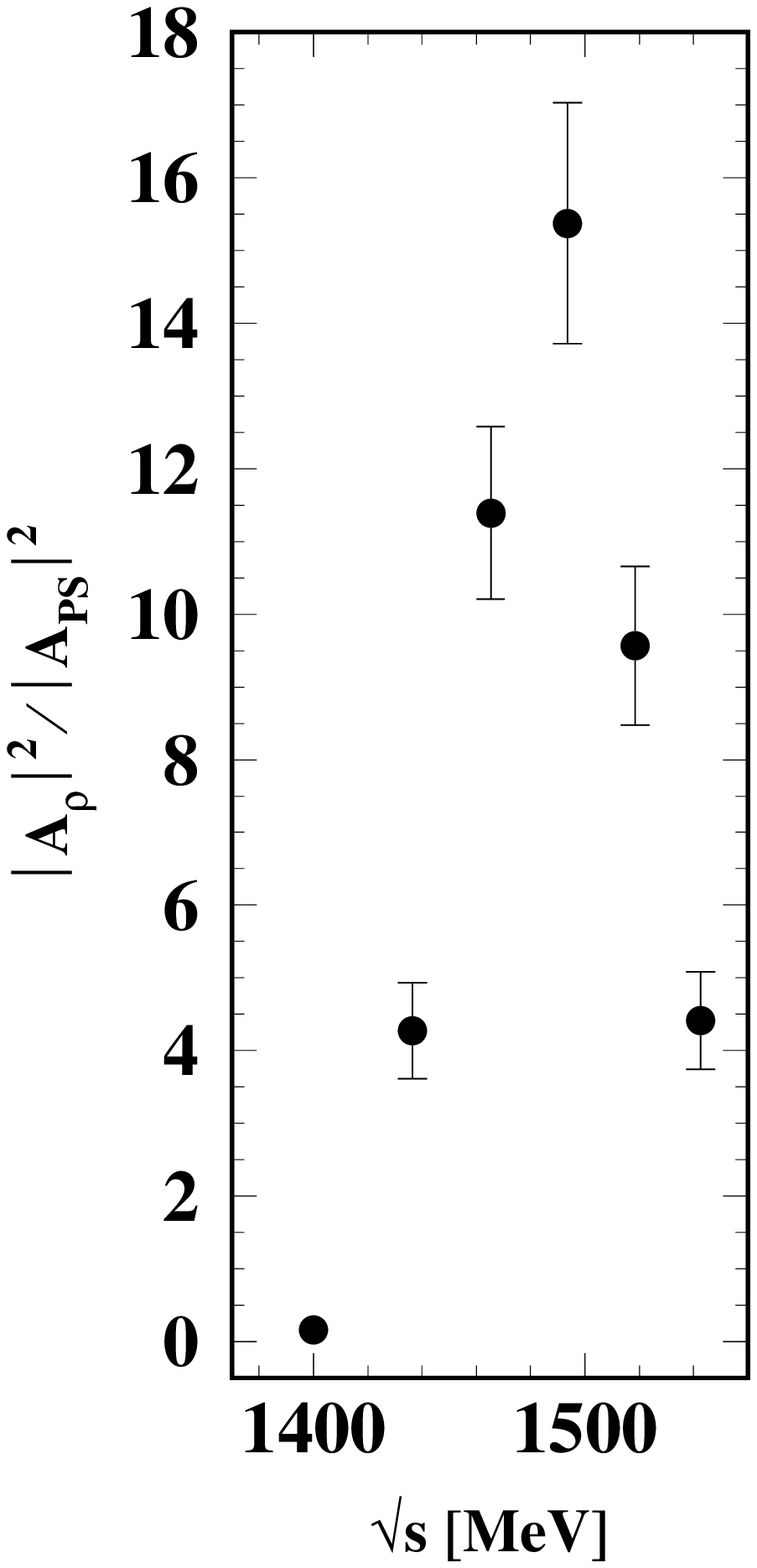}\\
\end{minipage}
\vspace*{-0.8cm}
\begin{center} \parbox{14cm}{
 \caption[]{\label{fig:3}
Invariant mass distribution of the pion pairs from the $\pi^+\pi^o$ final state
(full symbols). Left hand side: the measured
invariant mass distributions compared to the 
$\pi^o\pi^o$ final state (open symbols), to the expectation from phase space
behavior (dashed) and a fit including phase space component and
$\rho$-production. Right hand side: fitted ratio of matrix elements
for phase space and $\rho$-production.
}} 
 \end{center}
\end{figure}

Ochi et al. \cite{Ochi} suggested that a contribution of the 
$\rho$-Kroll-Rudermann term, which is negligible for the other isospin 
channels, might solve the problem. This suggestion motivated a carefull study 
of the invariant mass distributions of the pion - pion and the pion - nucleon 
pairs, which are again the most sensitive observables \cite{Langgaertner}. 
Contributions from $\rho^+$-meson production should result in an enhancement 
of high pion - pion invariant masses.

The DAPHNE collaboration \cite{Zabrodin} has searched for such enhancements in 
the $d(\gamma, \pi^o\pi^-)pp$ reaction and found some indication of the effect.
However, their analysis is largely complicated by effects from the bound 
nucleons. The pion - pion invariant masses for the $\pi^o\pi^o$ and 
$\pi^o\pi^+$ pairs from the free proton measured with TAPS 
\cite{Wolf,Langgaertner} are compared at the left hand side of 
fig.\ref{fig:3}. This comparison is particulary instructive since the 
$\rho^o\rightarrow\pi^o\pi^o$ decay is forbidden so that
the $\rho$-meson cannot contribute to the double $\pi^o$ channel. The 
$\pi^o\pi^o$ invariant mass is similar to phase space behavior, 
but at the higher incident photon energies the $\pi^o\pi^+$ invariant mass  
has an excess at large values.
The $\pi^o\pi^+$-data were fitted with a simple model assuming only phase space
and $\rho$-decay contributions \cite{Langgaertner}.
The result for the ratio of the matrix elements without phase space factors 
is shown at the right hand side of fig.\ref{fig:3}. 
The relative contribution of the $\rho$-decay matrix element peaks close to the
position of the D$_{13}$ resonance which is a hint at a 
D$_{13}\rightarrow N\rho$ contribution. 
In view of the new experimental results the group of E. Oset has updated their 
model \cite{Nacher} and now correctly included the possible $\rho$ 
diagrams. They found indeed a significant contribution of the
D$_{13}\rightarrow\ N\rho$ decay.

\begin{figure}[t]
\begin{minipage}{0.cm}
   \includegraphics[width=6.5cm]{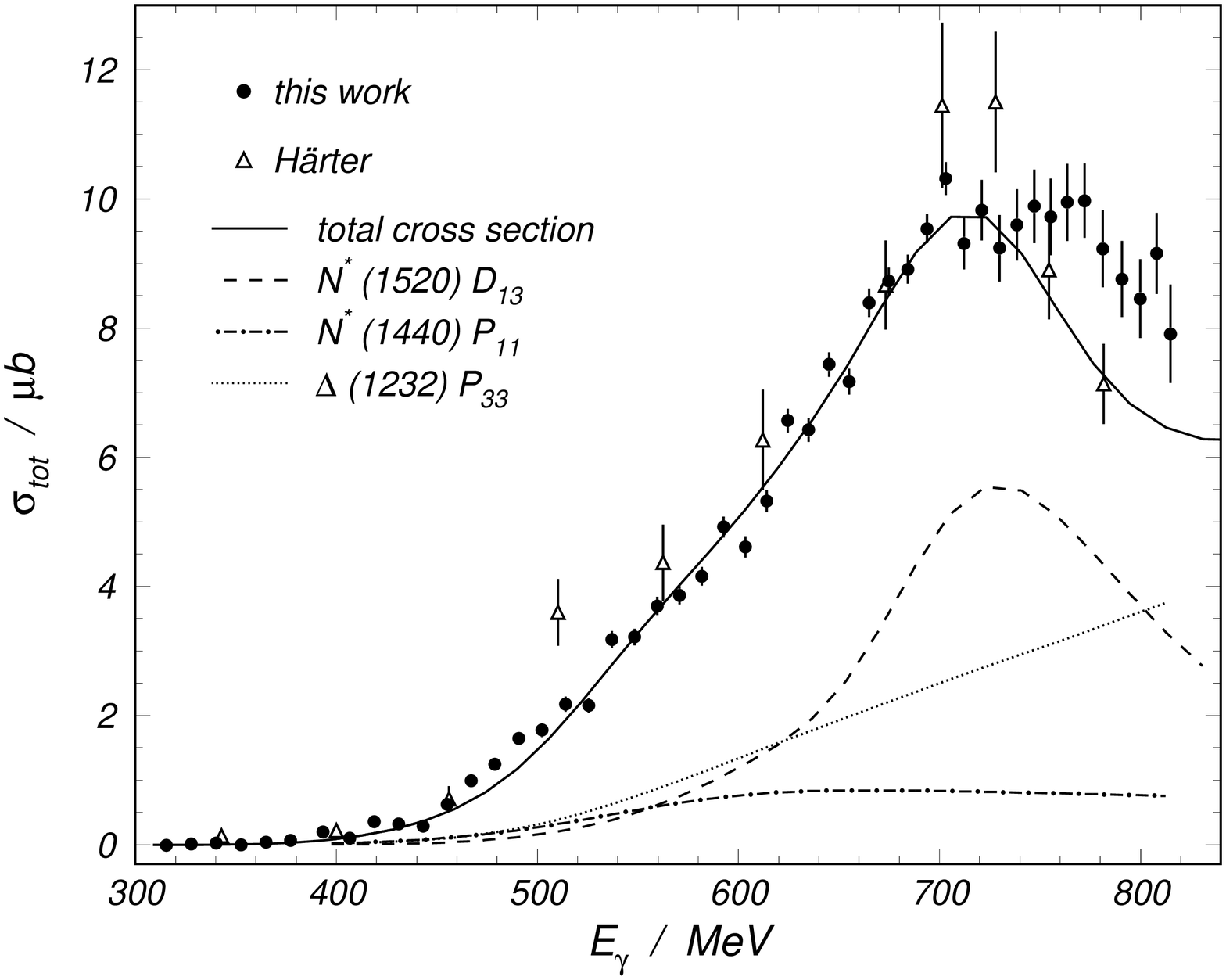}\\
\end{minipage}
\hspace{7.cm}
\begin{minipage}{0.cm}
\vspace*{-0.5cm}
   \includegraphics[width=6.5cm]{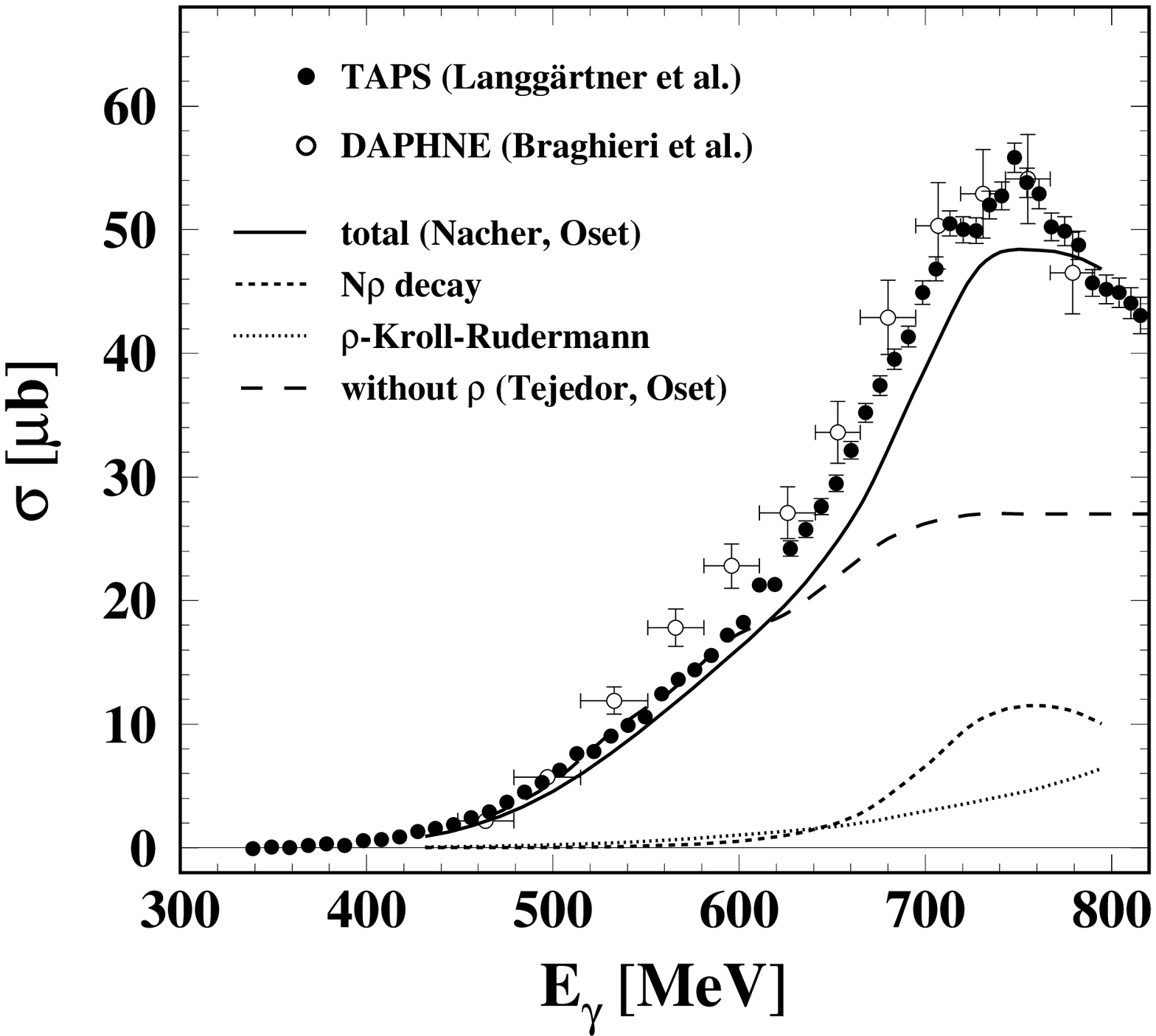}\\
\end{minipage}
\vspace*{-0.8cm}
\begin{center} \parbox{14cm}{
 \caption[]{\label{fig:0}
left side: total cross section for the $\gamma p\rightarrow p\pi^o\pi^o$
reaction compared to the results of the model from ref. \cite{Gomez}.
Right side: total cross section for the $\gamma p\rightarrow p\pi^o\pi^+$
reaction compared to the results with and without $\rho$-contributions
\cite{Gomez,Nacher}.
}} 
 \end{center}
\end{figure}

The final results are summerized in fig.\ref{fig:0} where the measured
total cross sections are compared to the latest model results. 
The prominant role of the sequential decay of the D$_{13}$ resonance via 
D$_{13}\rightarrow\Delta\pi\rightarrow N\pi^o\pi^o$ to the $\pi^o\pi^o$ final
state explains the peak at roughly 700 MeV incident photon energy. 
The peak like structure for the $\pi^o\pi^+$ final state is mainly caused 
by the D$_{13}\rightarrow N\rho$ decay which is enhanced via interference 
effects with other diagrams. It was shown earlier \cite{Gomez} that the 
peaking of the $\pi^+\pi^-$ final state comes from an interference of the  
D$_{13}\rightarrow\Delta\pi\rightarrow N\pi^+\pi^-$ with the leading $\Delta$-KR
term. These results taken together show that the resonance bump in double pion 
production is mainly due to the excitation of the D$_{13}$ resonance.

\section{The D$_{13}$(1520) excited in nuclei}

The in-medium properties of mesons and nucleon resonances are a very hotly
debated subject, however with the exception of the $\Delta$-isobar 
experimental results are still very scarce and partly contradicting.
We have therefore now investigated the D$_{13}(1520)$ resonance in the nuclear 
medium via quasifree single $\pi^o$-photoproduction.

\begin{figure}[t]
\begin{minipage}{0.cm}
   \includegraphics[width=6.5cm]{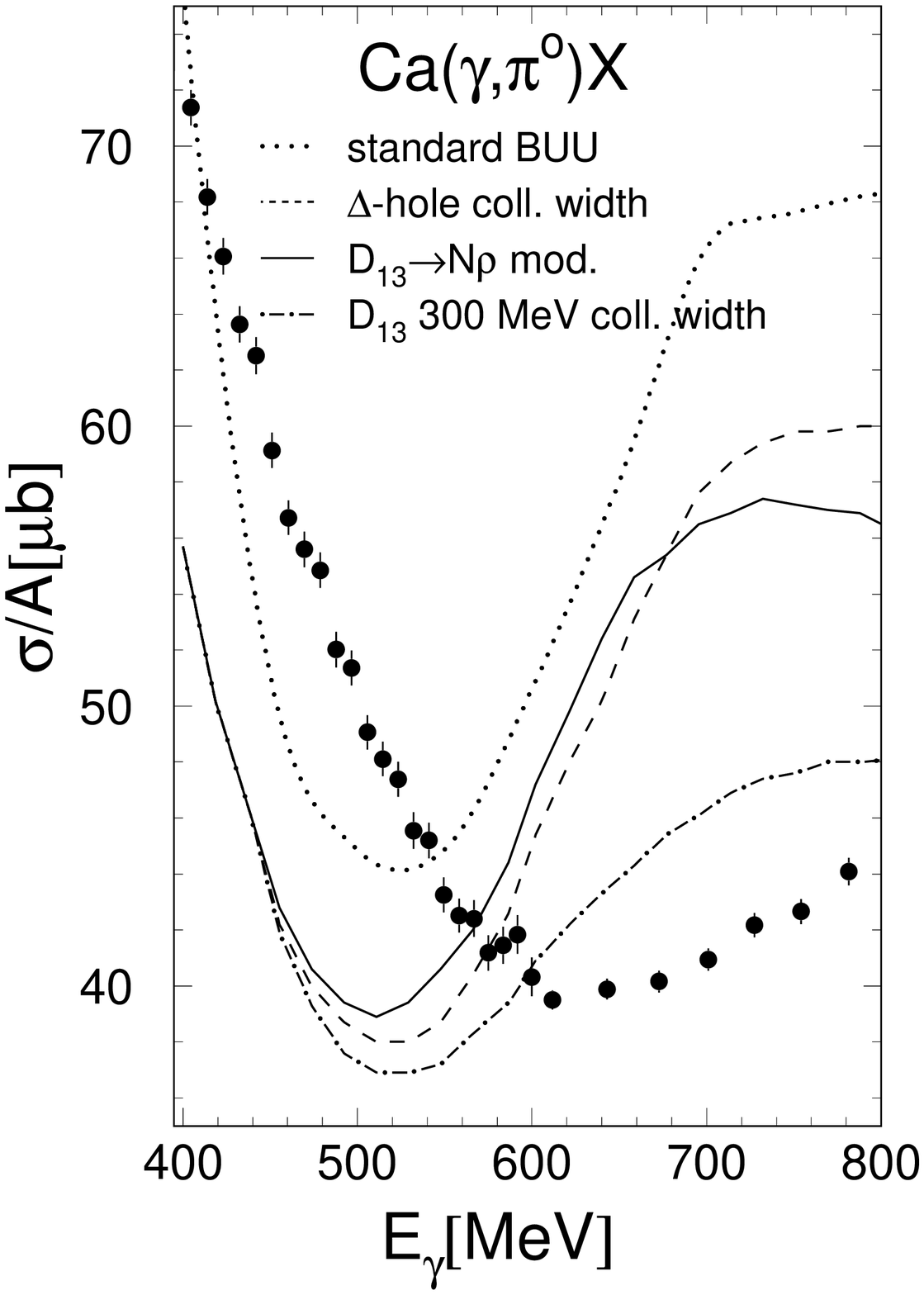}\\
\end{minipage}
\hspace{7.cm}
\begin{minipage}{0.cm}
   \includegraphics[width=5.8cm]{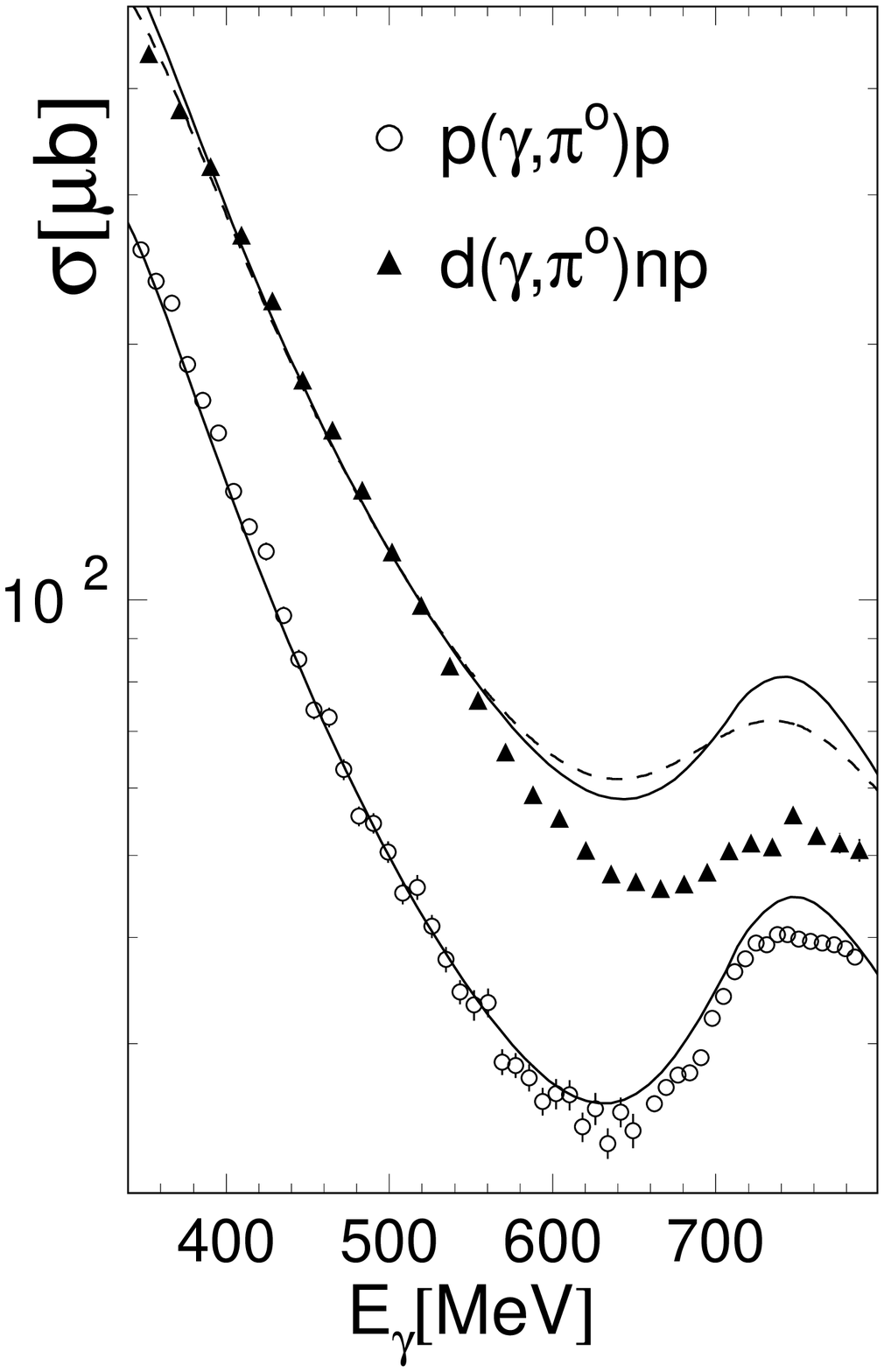}\\
\end{minipage}
\vspace*{-0.8cm}
\begin{center} \parbox{14cm}{
 \caption[]{\label{fig:4}
Left hand side: total inclusive cross section for Ca$(\gamma ,\pi^o)$X 
compared to BUU-model predictions \cite{Lehr}. Curves: standard BUU (dotted); 
BUU, $\Delta$-collisional width from $\Delta$-hole model (dashed); like dashed
but modified D$_{13}\rightarrow N\rho$-decay (solid), like dashed
but 300 MeV collisional width of D$_{13}$ (dash-dotted).  
Right hand side: quasifree single $\pi^o$-production from proton and
deuteron. Open symbols:
proton data, solid curve: MAID proton cross section; filled triangles: deuteron
data, solid curve: incoherent sum of MAID proton, neutron cross sections, 
dashed: like solid but Fermi smeared.
}} 
 \end{center}
\end{figure}

For the discussion of possible medium modifications of the D$_{13}$ resonance we
compare first the inclusive data to predictions in the framework of a transport
model (BUU) \cite{Lehr}. 
Here, inclusive means that all events with at least one $\pi^o$ are included.
The total cross section (left hand side of fig.\ref{fig:4}) predicted by the 
standard BUU calculation largely overestimates the data. Some improvement 
is achieved when the spreading width of the $\Delta$ is taken from $\Delta$-hole
models. However, even the calculation taking into account the
in-medium modification of the D$_{13}\rightarrow N\rho$ decay shows a much
larger resonance bump than is observed in the data. 
Only an arbitrary and probably unrealistic broadening \cite{Lehr} of the 
D$_{13}$-resonance by 300 MeV produces a significant suppression.

In the next step quasifree single $\pi^o$-photoproduction was selected via a
missing energy analysis as in \cite{Krusche_4}. The result for the deuteron is
compared at the right hand side of fig.\ref{fig:4} to the proton data and to
the expectation from a unitary isobar analysis of pion photoproduction (MAID) 
\cite{MAID}.
\begin{figure}[t]
\begin{minipage}{0.cm}
   \includegraphics[width=5.8cm]{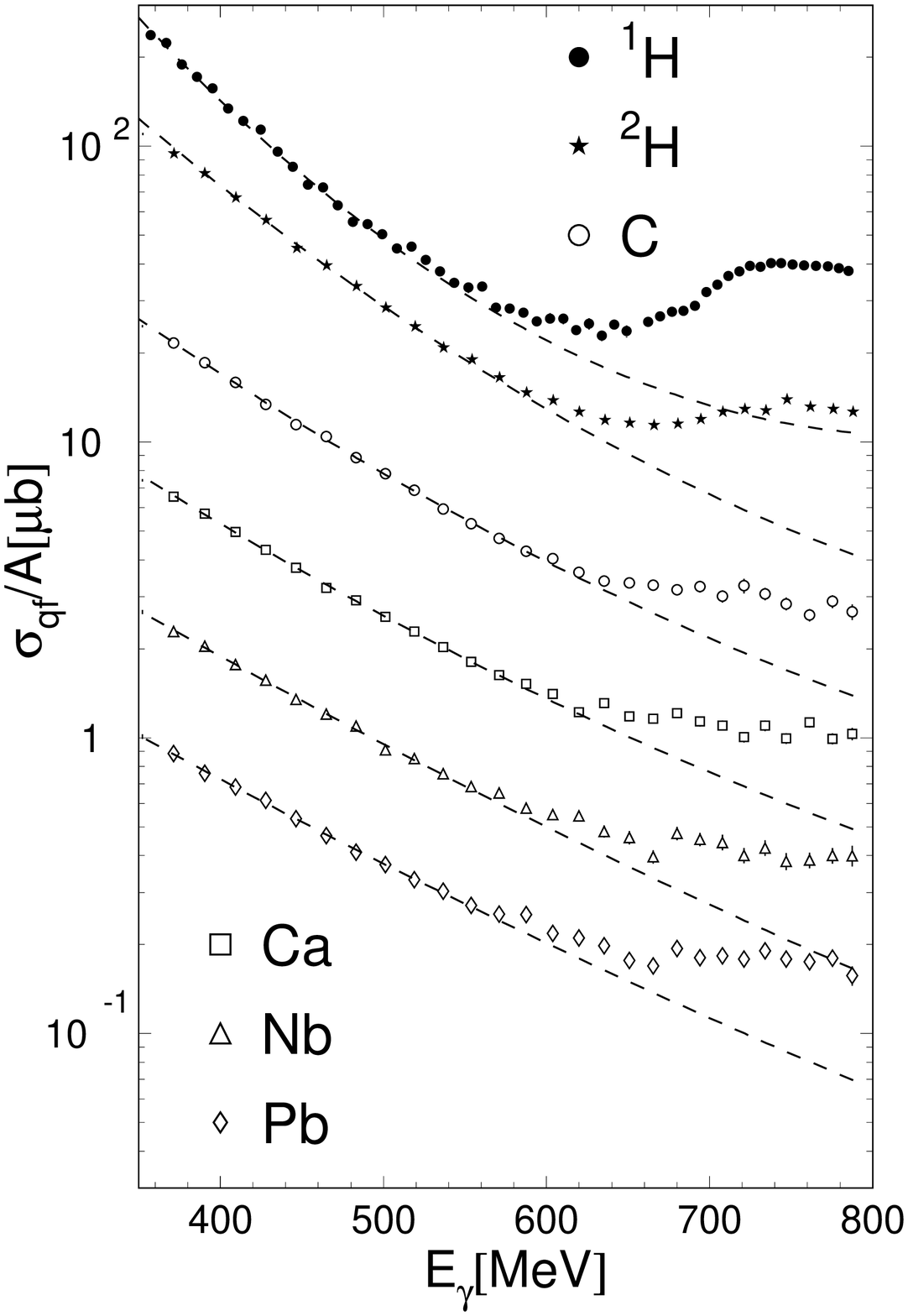}\\
\end{minipage}
\hspace{7.cm}
\begin{minipage}{0.cm}
   \includegraphics[width=5.8cm]{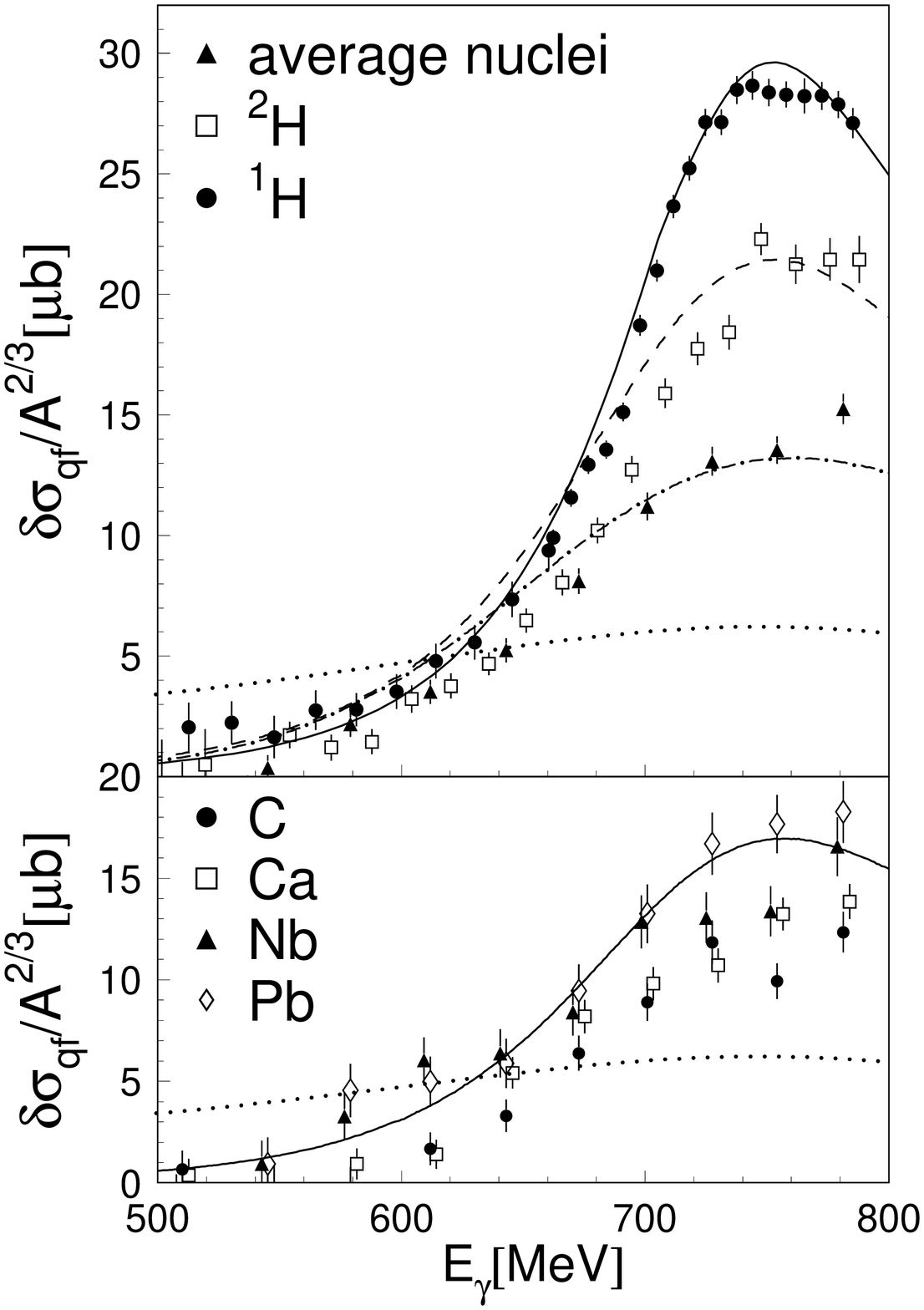}\\
\end{minipage}
\vspace*{-0.8cm}
\begin{center} \parbox{14cm}{
 \caption[]{\label{fig:5}
Left hand side: Total cross section per nucleon for single $\pi^o$ 
production for the nucleon and for nuclei.
Scale corresponds to proton, other data scaled down by factors 2,4,8,16,32. 
Dashed curves: fits of the data in the range 350 - 550 MeV.
Right hand side: (upper part) differences data - fits scaled by $A^{2/3}$. 
Full curve: MAID prediction for excitation of D$_{13}$ and S$_{11}$ on the 
proton; dashed curve: MAID proton - neutron average folded with deuteron Fermi 
momentum; dash-dotted: folded with Fermi momentum for medium weight nuclei
(both scaled to the data).
Dotted curve: Breit-Wigner curve for the D$_{13}$ resonance with 300 MeV width.
lower part: individual nuclear data and prediction from deuteron cross section
(solid curve, see text).
}} 
 \end{center}
\end{figure}

The data for the proton are well reproduced. The deuteron data are compared
to the sum of the proton and neutron cross sections from MAID folded with the
momentum distribution of the bound nucleons (see fig.\ref{fig:4}, right side).
The result agrees very well with the data in the tail of the 
$\Delta$-resonance, but it largely overestimates the data in the D$_{13}$ 
region. This result is surprising since we are dealing with {\it quasifree} 
pion production, for which the large momentum mismatch between participant and 
spectator nucleon is expected to suppress any interference terms between the 
two nucleons. This finding could be part of the explanation why the BUU 
results overestimate the inclusive data. Such calculations must rely on the 
assumption that the total cross section from nuclei before taking into account 
in-medium and FSI effects is the incoherent sum of proton and neutron cross 
sections. 

For a more quantitative analysis of the D$_{13}$-excitation in nuclei the
cross sections were decomposed into a resonance and a background part. In
principle, such a decomposition requires a multipole analysis wich takes into
account resonance - background interference terms. However, interference terms 
are small in this case \cite{Krusche_5}.
The result is shown in fig.\ref{fig:5} (left hand side). 
The background part was fitted with a function of the type:
\begin{equation}
\sigma\propto e^{(aE_{\gamma}^2+bE_{\gamma})} 
\end{equation}
with $a$ and $b$ as free parameters. 
The resonance contribution for heavier nuclei is not qualitatively 
different from the deuteron case.
The differences between measured cross sections and fits are
shown in fig.\ref{fig:5} (right hand side).
In the upper part of the figure the resonance contributions for the proton,
the deuteron and the average for the nuclei are compared to the MAID
predictions for the D$_{13}$ and S$_{11}$ contributions folded with the Fermi
momentum distributions and scaled to the data. No broadening of the
resonance structure beyond Fermi smearing is observed. A D$_{13}$ resonance
broadened to 300 MeV as used in the BUU calculations \cite{Lehr} for the 
inclusive data (see fig.\ref{fig:4}) is clearly ruled out, the data correspond
rather to BW-curves with a width around 100 MeV.

Finally, it was investigated if the strength of the resonance signal 
for the nuclei is consistent with the deuteron case.  
The MAID proton cross section for resonance excitation was folded with the 
nucleon momentum distribution and compared to the measured deuteron cross 
section. Agreement is obtained for 
$\sigma_n(\mbox{D}_{13})/\sigma_p(\mbox{D}_{13})\approx 1/3$. We have then 
adopted the 1/3 ratio, folded $(1+1/3)\sigma_p/2$ with a
typical nuclear momentum distribution and compared the result to the
nuclear data scaled to $A^{2/3}$, which in first approximation accounts 
for the FSI effects (see fig.\ref{fig:5}, lower part). The agreement of this 
approximation with the data is quite good.

The approximate scaling of the cross sections with $A^{2/3}$ indicates 
of course FSI effects. This means that in contrast to total photoabsorption 
not the entire nuclear volume  is probed. 
However, suppression of the resonance bump in total photoabsorption reactions 
occurs already for $^4$He \cite{MacCormick} and
does not change from very light nuclei like lithium and beryllium up to very 
heavy ones like uranium. This excludes a strong density dependence of the 
effect.

Furthermore, the models without a strong broadening of the D$_{13}$ resonance 
overestimate our inclusive pion data (see fig.\ref{fig:4}) which are subject 
to FSI in a similar way as the exclusive data. 
It thus seems that the models miss some other effect which must be understood 
before the results from total photoabsorption can be used as evidence for an 
in-medium resonance broadening. 

In summary, investigating quasifree $\pi^o$ photoproduction from nuclei we have
found no indication of a broadening or a depletion of the excitation strength
of the D$_{13}(1520)$ resonance compared to the deuteron.
On the other hand, in comparison to the free proton, the resonance structure 
for the deuteron in the D$_{13}$-region is much reduced, but not broadened.

\end{document}